\documentclass{article}
\usepackage[utf8]{inputenc}
\usepackage[english]{babel}
\usepackage{xcolor}
\usepackage{hyperref}
\usepackage{cite}  
\usepackage{algorithm} 
\usepackage{algpseudocode} 
\usepackage{graphicx}
\usepackage{amsmath}
\usepackage{bm}
\usepackage{soul}
\usepackage{authblk}

\title{A Geometric Approach for Computing the Kernel of a Polyhedron}
\author{T.~Sorgente}
\author{S.~Biasotti}
\author{M.~Spagnuolo}
\affil{IMATI, Consiglio Nazionale delle Ricerche, Genova, Italy}

\begin{document}
\maketitle
 
\begin{figure}
    \centering
    \includegraphics[width=\linewidth]{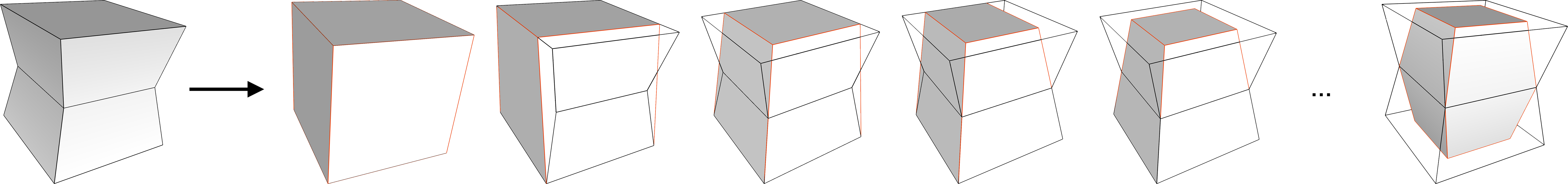}
    \caption{Pipeline of the kernel computation for a polyhedron. At first step, we compute the Axis Aligned Bounding Box (AABB) of the polyhedron; then, we iterate on each face $f$ of the polyhedron (black edges) and cut AABB with the plane induced by $f$ (red edges).}
    \label{fig:pipeline}
\end{figure}

\begin{abstract}
   We present a geometric algorithm to compute the geometric \textit{kernel} of a generic polyhedron. 
   The geometric kernel (or simply kernel) is defined as the set of points from which the whole polyhedron is visible. 
   Whilst the computation of the kernel for a polygon has already been largely addressed in the literature, less has been done for polyhedra.
   Currently, the principal implementation of the kernel estimation is based on the solution of a linear programming problem.
   We compare against it on several examples, showing that our method is more efficient in analysing the elements of a generic tessellation.
   Details on the technical implementation and discussions on pros and cons of the method are also provided.
\end{abstract}  

\section{Introduction}
\label{sec:intro}

The concept of \emph{geometric kernel} of a polygon, a polyhedron, or more generally of a shape, is a pillar of computational geometry. 
Intuitively, the kernel of a close, geometric shape $S$ is the locus of the points internal to $S$ from which the whole shape $S$ is visible.
This concept is particularly interesting when applied to non convex polytopes, as for convex shapes the kernel coincides with the shape itself. 

In the simplest scenario, that is if the shape is a polygon, the standard way of computing its kernel is by intersecting appropriate half-planes generated from its edges.
This problem has been tackled since 70s, when \cite{ShamosHoey} presented an efficient algorithm  that performed the kernel computation in $O(e\log e)$ operations, where $e$ is the number of edges of a polygon, as the intersection of $e$ half-edges.
After that, an optimal algorithm able to run in $O(n)$ operations, being $n$ the number of vertices of the polygon, has been proposed in \cite{LeePreparata}. 
Up to our knowledge, computational tools and libraries like \textit{Boost} \cite{BoostLibrary}, \textit{Geogram} \cite{levy2015geogram}, \textit{CGAL} \cite{fabri2009cgal}, or \textit{Libigl} \cite{jacobson2017libigl} implement routines to compute intersections between polygons and planes, which can be used to estimate the kernel.
In the first attempts, for example in \cite{PreparataShamos}, the extension of the problem to the 3D case was treated only from a theoretical point of view.
Starting from this perspective, the natural approach of extending of the 2D method (which we call the \textit{geometric approach)} was soon dismissed as unattractive for computational reasons.
It was replaced by a new approach (which we call \textit{algebraic)} which makes use of linear algebra and homogeneous coordinates, and that is the state of the art for computing 3D kernels currently implemented by libraries like CGAL.

During years, the polygon kernel computation has become popular to address several problems based on simple polygon analysis, such as star-component decomposition and visibility algorithms that are of interest in robotics, surveillance, geometric modeling, computer vision and, recently, in the emerging field of additive manufacturing \cite{demir2018near}.
Today, the geometric kernel of a polytope is a pivotal information for understanding the geometrical quality of an element in the context of finite elements analysis.
While in the past years finite elements methods were only designed to work on convex elements like triangles/tetrahedra or quadrangles/hexahedra  \cite{ciarlet2002finite}, recent and more complex methods like the Mimetic Finite Difference Method \cite{lipnikov2014mimetic}, the Virtual Elements Method \cite{beirao2013basic}, the Discontinuous Galerkin Mehod \cite{cockburn2012discontinuous} or the Hybrid High Order Method \cite{di2019hybrid} are able to deal with non convex polytopes.
This enrichment of the class of admissible elements led researchers to further investigate the idea of the geometric quality of a polytope, and to define quality measures and metrics for the elements of a mesh \cite{attene2021benchmark,sorgente2021role}.
In this setting, the geometric kernel is often associated with the concepts of \textit{shape regularity} and \textit{star-shapedness} of an element.
For example, as analyzed in \cite{sorgente2021vem}, most of the error estimates regarding the VEM (but the same holds for other polytopal methods) are based on the theory of polynomial approximation in Sobolev spaces, assuming the star-shapedness of the elements \cite{Brenner-Scott:2008,dupont1980polynomial}.
As a consequence there are a number of sufficient geometrical assumptions on the computational domain for the convergence of the method, which require an estimate of the kernel. 
Mesh quality measures/metrics/indicators require to compute the kernel of thousands of polytopes, each of them with a limited number of faces and vertices, in the shortest possible time \cite{sorgente2021role}.

In this paper we define an algorithm for the implementation of the geometric approach to the computation of the kernel of a polyhedron, and empirically show how this approach can significantly outperform the algebraic one when applied in the context of finite elements methods, with elements having a limited number of faces and vertices.

The paper is organized as follows.
In Section~\ref{sec:preliminary} we introduce and define the concept of kernel of a polytope in dimension 2 and 3.
In Section~\ref{sec:kernel} we detail the algorithm for the construction of the kernel of a polyhedra.
In Section~\ref{sec:examples} we exhibit some examples of computed kernels and analyze the performance of the algorithm, also with comparisons with an implementation of the algebraic approach.
In Section~\ref{sec:conclusions} we sum up pros and cons of the algorithm and draw some conclusions.

\section{Terminology and preliminary concepts}
\label{sec:preliminary}
Let us introduce some basic concepts, useful to the computation of the kernel of a polyhedron.
Following the notation adopted in \cite{PreparataShamos}, a \textit{polyhedron} is defined by a finite set of plane polygons such that every edge of a polygon is shared by exactly one other polygon and no subset of polygons has the same property.
The vertices and the edges of the polygons are the \textit{vertices} and the \textit{edges} of the polyhedron; the polygons are the \textit{faces} of the polyhedron.
In this work we only consider \textit{simple} polyhedra, which means that there is no pair of nonadjacent faces sharing a point.

A polyhedron $P$ is \textit{convex} if, for any two points $p_1$ and $p_2$ in $P$, the segment $(p_1,p_2)$ is entirely contained in $P$.
It can be shown that the intersection of convex polyhedra is a convex polyhedron.
Two points $p_1$ and $p_2$ inside $P$ are said to be \textit{visible} if the line segment connecting $p_1$ and $p_2$ does not intersect with the exterior of the $P$. 
It is easily seen that any two points inside a convex polyhedron are visible.
The \textit{kernel} of a $P$ is the set of points from which all points inside $P$ are visible. 
Some polyhedra may not have a kernel, or we also say their kernel is empty.
The polyhedron $P$ is called \textit{star-shaped} if there exists a sphere completely contained in its kernel.
A polyhedron is not star-shaped if its kernel is empty.

\begin{figure}[htbp]
\centering
    \includegraphics[width=\linewidth]{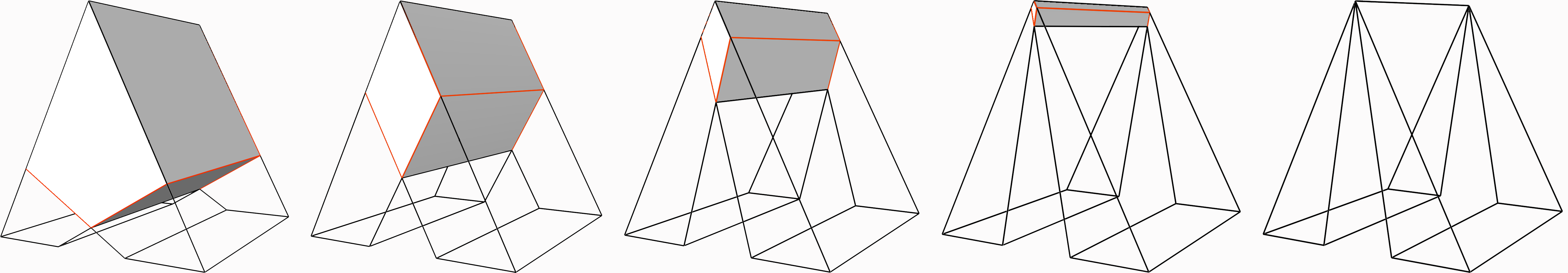}
\caption{Evolution of the kernel (in red) in a sequence of parametric polyhedra.}
\label{fig:tent}
\end{figure}

In Fig.~\ref{fig:tent} we present a parametric polyhedron shaped like a tent, with the parameter regulating the height of the "entrance".
As the parameter increases, the set of points from which the whole polyhedron is visible becomes smaller, and so does the kernel.
The last example of Fig.~\ref{fig:tent} is not star-shaped anymore, i.e. the kernel is empty.

In Section~\ref{sec:kernel} we will make use of the concept of \textit{signed distance}.
Given a plane $p$ with unit normal vector $\boldsymbol n$ and a point $\boldsymbol p$ on it, the signed distance between a point $\boldsymbol x$ and the plane is given by the scalar product $d(\boldsymbol x, p) = \boldsymbol n\cdot(\boldsymbol p-\boldsymbol x)$.
We say that $\boldsymbol x$ is strictly (weakly) below $p$ if $d(\boldsymbol x, p)<0$ $(\le0)$, $\boldsymbol x$ is strictly (weakly) above if $d(\boldsymbol x, p)>0$ $(\ge0)$, and $\boldsymbol x$ is on $p$ if $d(\boldsymbol x, p)=0$.


\subsection{Geometric vs algebraic approach}
\label{subsec:geometric_algebraic}
The state of the art algorithm in the 2D case follows a geometric approach: the kernel of a polygon is found as the intersection of half-planes originating from its edges.
We use the term "geometrical" because the algorithm computes repeatedly a sequence of geometric intersections between polygons and planes.
This idea was optimized until obtaining an algorithm able to run in $O(n)$ operations, which has been proven to be optimal.
One natural way to define a method for the 3D kernel computation is to extend the 2D algorithm, which is well studied and documented, to the upper dimension.

The problem with the 3D case is that whereas two convex polygons with respectively $n_1$ and $n_2$ vertices can be intersected in time $O(n)$, being $n=n_1+n_2$, two convex polyhedra with the same parameters are intersected in time $O(n\log n)$, thus the generalization of the two-dimensional instance would yield an $O(n\log^2 n)$ algorithm.
This is in contrast with the result shown in \cite{PreparataShamos}, where a lower bound for the intersection of convex polyhedra is established at $O(n\log n)$.

This brought to the definition of a new algorithm based on the so-called "double duality trick", which makes use of linear algebra and homogeneous coordinates, able to compute the intersection of $n$ half-spaces in time $O(n\log n)$ \cite{PreparataShamos}.
This algebraic approach can be implemented inside the framework of the CGAL library, although there is currently not an explicit routine for computing the kernel of a polyhedron and one has to connect the function for the intersection of half-spaces to the polyhedron data structure.

While from a theoretical point of view the cited results are indubitable, we believe that in many practical situations the geometric approach could perform better than the algebraic one.
Intuitively, if the number of faces and vertices of the polyhedron is low, this method can be more efficient than the algebraic one, as the cost of solving a linear problem does not go under a certain bound while the intersection of half-spaces can become extremely cheap if performed intelligently.

\section{Computing the kernel of a Polyhedron}
\label{sec:kernel}
In this section, we induce our method for computing the kernel of a polyhedron.
It has a modular structure composed of four nested algorithms, each one calling the next one in its core part.
It is modular in the sense that each algorithm can be entirely replaced by another one performing the same operation(s).
For instance, one could use another strategy for computing the intersection between a polygon and a plane and simply replace Algorithm~\ref{alg:polygon-plane} (and Algorithm~\ref{alg:line-plane} if not needed).

In the next subsections we adopt the following data structure inherited by the \textit{cinolib} library \cite{livesu2019cinolib}, in which the code has been written:
\begin{itemize}
    \item \textit{Polyhedron:} class composed by a field \textit{verts} containing the vertices (in 3D coordinates) and a field \textit{faces} containing the faces of a polyhedron.
    \item \textit{Points:} array of unordered 3D points.
    \item \textit{Face:} array of unsigned integers representing the indices of the vertices of a face, ordered counter-clockwise.
    \item \textit{Plane:} class composed by a 3D point $d$ indicating a random point on the plane, and a 3D point $n$ indicating the unit normal of the plane.
\end{itemize}
We point out that we always consider a plane $p$ together with the direction indicated by its normal vector $p.n$, which is equivalent to considering the half-space originating in $p$ and containing $p.n$.


\subsection{Polyhedron Kernel}
\label{subsec:polyhedron_kernel}
Algorithm~\ref{alg:kernel} tackles the main problem: given a polyhedron $P$, we want to find the polyhedron $K$ representing the kernel of $P$.
We will also need as input an array containing the outwards normals of the polyhedron faces, as it is not always possible to compute the orientation of a face only from its vertices (for example with non-convex faces).

We start by defining $K$ as the axis aligned bounding box (AABB) of $P$, i.e. the box with the smallest volume within which all the vertices of $P$ lie, aligned with the axes of the coordinate system.
We then recursively "cut" this box with a number of planes.
For each face $f$ of $P$ we compute the plane $p$ which contains it, with the orientation given by the opposite of its normal $N(f)$ (that is to say, $p.n=-N(f)$).
In general $p$ will separate $K$ into two polyhedra, and between those two we choose the one containing the vector $p.n$, which points towards the interior of $K$.
This operation is performed by the \textit{Polyhedron-Plane-Intersection} algorithm detailed in Section \ref{subsec:polygon_plane}, which replaces $K$ with the chosen polyhedron.

We point out that cutting a convex polyhedron with a plane will always generate two convex polyhedra, and since we start from the bounding box (which is convex), we are guaranteed for $K$ to be always a convex polyhedron.
We could as well start with considering the polyhedron's convex hull instead of its bounding box, but it would be less efficient because the convex hull costs in general $O(n\log n)$ while the AABB is $O(n)$.
Note that even if we used the convex hull, we would still need to intersect the polyhedron with each of its faces.
As we iterate through the faces, we generate a sequence of convex polyhedra $K_i$, $i=1,\dots,\#$\textit{faces}, such that $K_i\subseteq K_{i-1}$.
No matter how weird the initial element $P$ is, from this point on we will only be dealing with convex polyhedra and convex faces.

\begin{algorithm}[htbp]
\caption{Polyhedron-Kernel}
\label{alg:kernel}
\begin{algorithmic}[1]
\Require Polyhedron $P$, Points $N$ (faces normals);
\Ensure Polyhedron $K$
\State $K$ := AABB of $P$;
\For{Face $f$ in $P.faces$}
    \State Plane $p$ := plane containing $f$ with normal $-N(f)$;
    \State $K$ := Polyhedron-Plane-Intersection($K$, $p$);
\EndFor
\State \Return $K$;
\end{algorithmic}
\end{algorithm}


\subsection{Polyhedron-Plane-Intersection}
\label{subsec:polyhedron_plane}
With the second algorithm we want to intersect a polyhedron $P$ with a plane $p$.
This intersection will in general be composed of two polyhedra, and between these two we are interested in the one containing the normal vector of $p$ (conventionally called the one "above" the plane and indicated with $A$).
This algorithm is inspired from \cite{ahn2008geometric}, where the authors define an algorithm for the intersection of a convex polyhedron with an half-space.

\begin{figure}[htbp]
\centering
\begin{tabular}{cc}
\includegraphics[width=.49\linewidth]{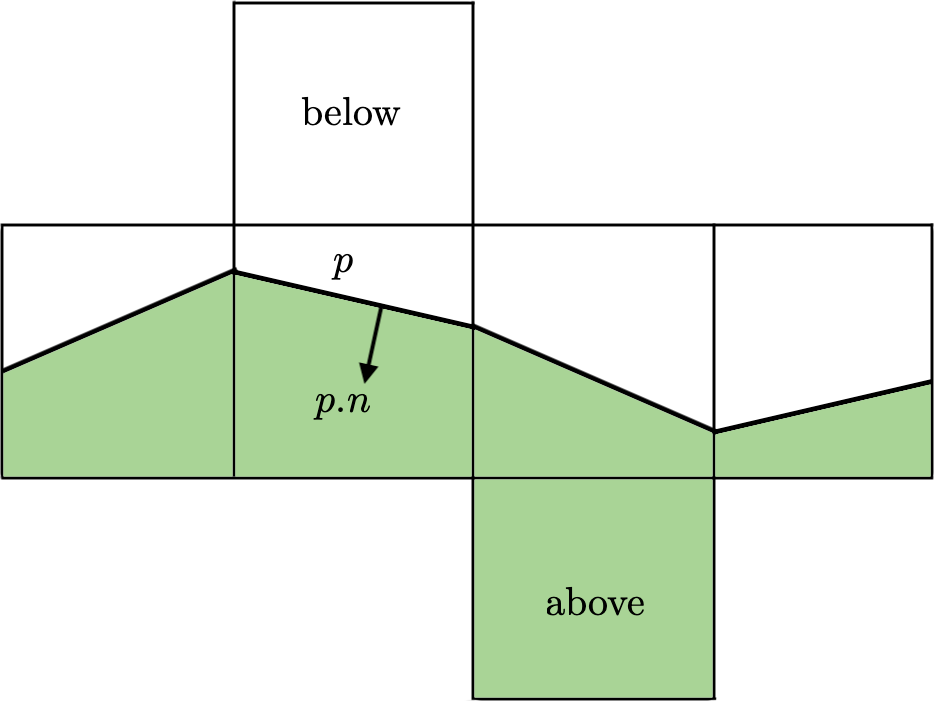} &
\includegraphics[width=.49\linewidth]{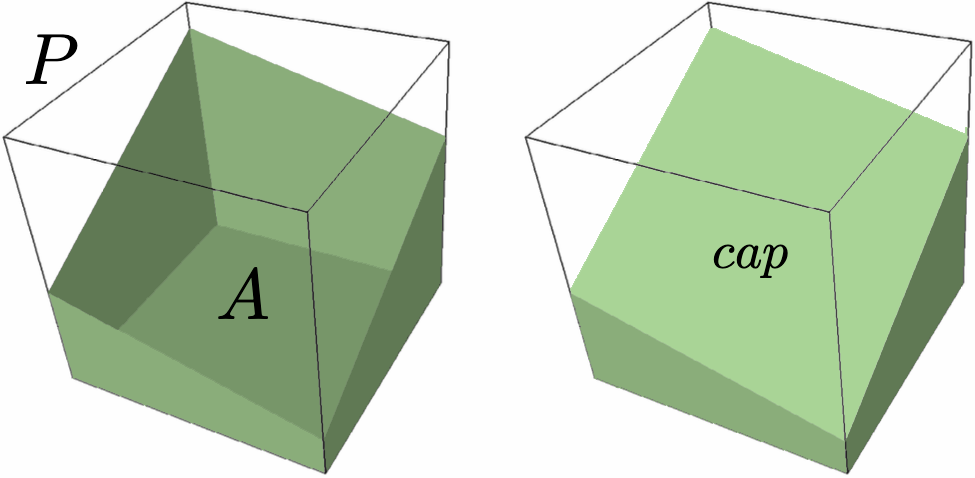}\\
(a) & (b)
\end{tabular}
\caption{Intersection of a polyhedron with a plane: (a) clipping and (b) capping of a cube.}
\label{fig:polyhedron-plane}
\end{figure}

The first part of Algorithm~\ref{alg:polyhedron-plane} is called the "clipping" part (recalling the terminology from \cite{ahn2008geometric}) and consists in cutting each face of $P$ with the plane $p$, see Fig.~\ref{fig:polyhedron-plane}(a).
We preliminarly evaluate the signed distances (defined in Section~\ref{sec:preliminary}) between the vertices of $P$ and the plane to understand their position with respect to $p$.
Faces strictly below $p$ are discarded, faces weakly above $p$ are added to $A$ together with their vertices, and faces intersected by $p$ are split by the \textit{Polygon-Plane-Intersection} algorithm.
While we visit every face only once, the same does not hold for vertices, hence we need to check if a vertex is already present in $A$ before adding it.

This simple idea of processing in advance the faces which are only weakly above the plane resolves several implementation issues and in some cases significantly improves the efficiency of the algorithm.
By doing this, we make sure that only the faces properly intersected by the plane are passed to Algorithm~\ref{alg:polygon-plane}, so that we do not need to implement all the particular cases of intersections in a single point or along an edge or of faces contained in the plane.
In addition, for every face not passed to Algorithm~\ref{alg:polygon-plane} we have an efficiency improvement, and this happens frequently in models with many coplanar faces like the ones considered in Section~\ref{subsec:refinements}.

If at the end of this step $p$ intersects $A$ in more than two points, given that $A$ and all its faces are convex, the vertices shared by $p$ and $A$ will define a "cap" face of $A$ completely contained in $p$, see Fig.~\ref{fig:polyhedron-plane}(b).
In order to sort counter-clockwise the points contained in \textit{capV} we project them onto a plane and apply the algorithm proposed in \cite{baeldung} for 2D points.
Note that if the cap face was not convex it would make no sense to order its vertices, but the intersection between a plane and a convex polyhedron will always generate convex faces.
Last, we need to check that this face is not already present in $A$: for example if $p$ is tangent to $P$ along a face, this face could be added to $A$ both as an intersection face and as a cap face.
If this is not the case we add \textit{capF} to $A.faces$, but we do not need to add any vertex from \textit{capV}, as we can assume they are all already present in $A.verts$.

\begin{algorithm}[htbp]
\caption{Polyhedron-Plane-Intersection}
\label{alg:polyhedron-plane}
\begin{algorithmic}[1]
\Require Polyhedron $P$, Plane $p$
\Ensure Polyhedron $A$
\State evaluate the position of all points in $P.verts$ with respect to $p$;
\For{Face $f$ in $P.faces$}
    \State Points $f_v=$ vertices in $P.verts$ relative to $f$;
    \If{all vertices in $f_v$ are strictly below $p$}
        continue;
    \ElsIf{all vertices in $f_v$ are weakly above $p$}
        \State $A.verts\leftarrow$ $f_v$, 
        $A.faces\leftarrow f$;
    \Else 
        \State \textit{aboveV},\textit{aboveF}:=Polygon-Plane-Intersection$(f_v,f,p)$;
        \State $A.verts\leftarrow$ \textit{aboveV},
        $A.faces\leftarrow$ \textit{aboveF};
    \EndIf
\EndFor
\State Points \textit{capV}:= vertices in $A.verts$ which are on $p$;
\If{size(\textit{capV})$<3$} return $A$;
\EndIf
\State Face \textit{capF} := \textit{capV} indices ordered counter-clockwise;
\If{\textit{capF} $\notin$ $A.faces$}
    $A.faces\leftarrow$ \textit{capF};
\EndIf
\State \Return $A$;
\end{algorithmic}
\end{algorithm}


\subsection{Polygon-Plane-Intersection}
\label{subsec:polygon_plane}
\begin{figure}[htbp]
\centering
\begin{tabular}{cc}
\includegraphics[width=.4\linewidth]{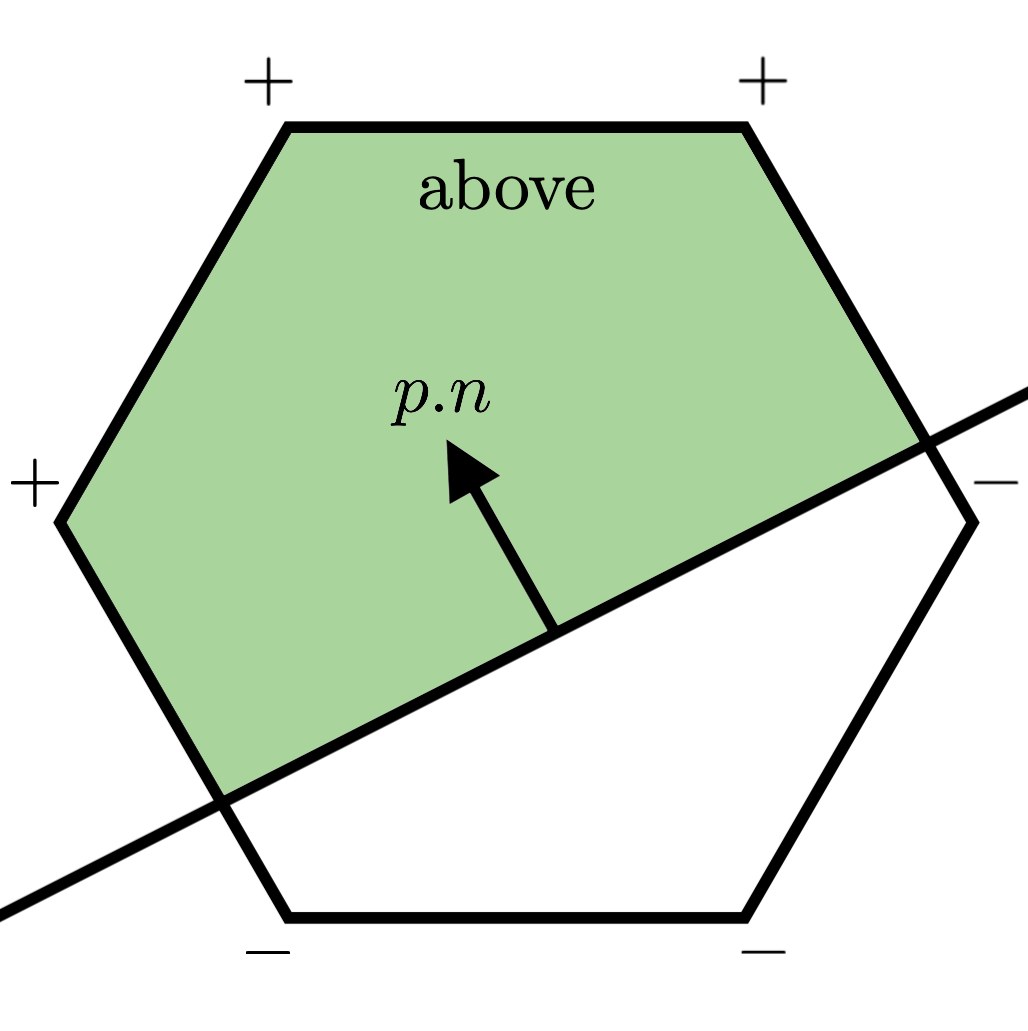} &
\includegraphics[width=.4\linewidth]{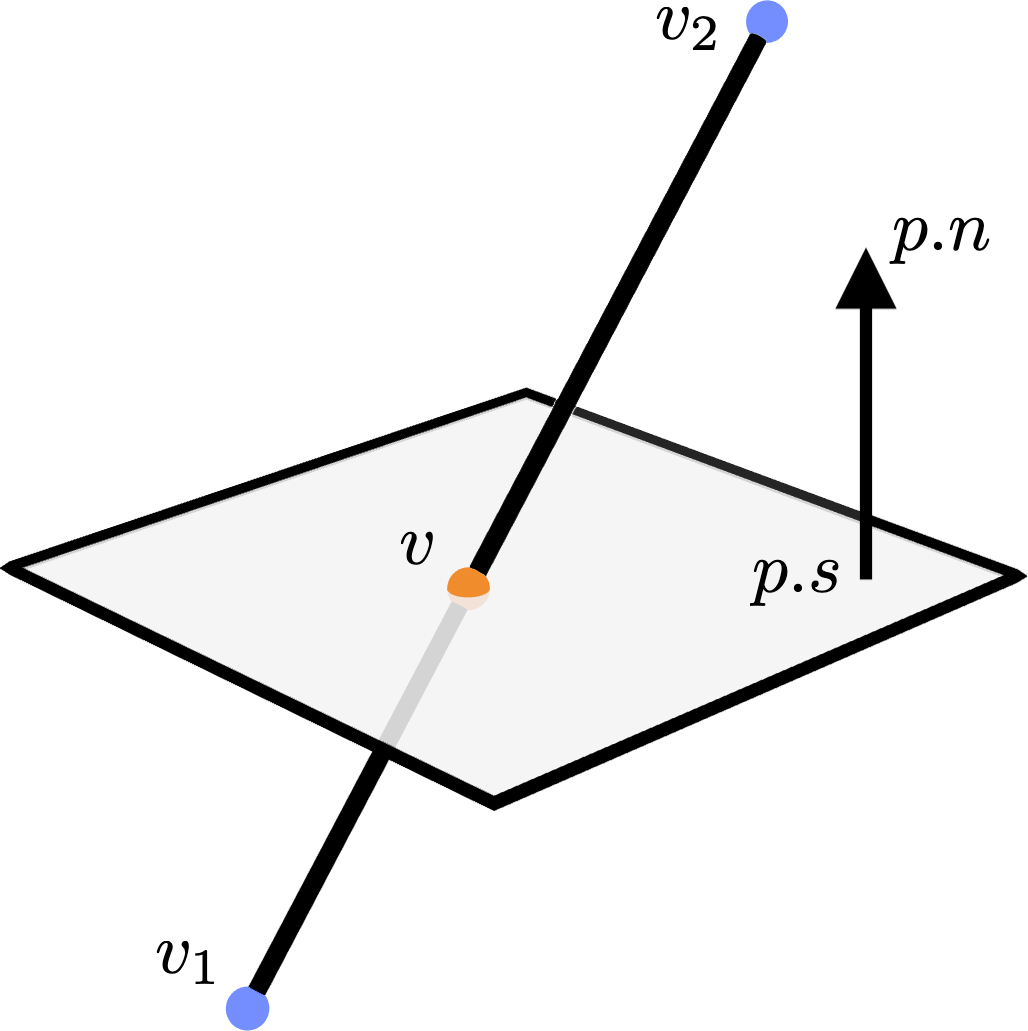}\\
(a) & (b)
\end{tabular}
\caption{(a) Intersection between a polygon and a plane, with the above part coloured in green. (b) Intersection between a line and a plane.}
\label{fig:polygon-line-plane}
\end{figure}

Algorithm~\ref{alg:polygon-plane} describes the intersection of a polygon (representing a face of the polyhedron), defined by an array of 3D points \textit{polyV} and an array of indices \textit{polyF}, with a plane $p$.
In analogy to Algorithm~\ref{alg:polyhedron-plane}, the intersection will in general produce two polygons and we are only interested in the one above the plane, see Fig.~\ref{fig:polygon-line-plane}(a), defined by vertices \textit{aboveV} and indexes \textit{aboveF}.
We generically say that a vertex $v$ is added to \textit{above} meaning that $v$ is added to \textit{aboveV} and its index $id_v$ is added to \textit{aboveF}.

We preliminarly evaluate the signed distances between the vertices of the polygon and the plane to understand their position with respect to $p$.
Then we iterate on the edges of \textit{poly}: in order to avoid duplicates, for each couple of consecutive vertices $v_1, v_2$, we only accept to add to \textit{above} the second vertex $v_2$ or the intersection vertex $v$, but never $v_1$.
When applying the algorithm recursively, as in the case of kernel computation, this idea requires to have all faces oriented coherently.

If both vertices are strictly below $p$ we ignore them, unless $v_2$ lies exactly on the plane, in which case we add it to \textit{above}.
If they are both above or on $p$ we add $v_2$ to \textit{above}, otherwise we perform the \textit{Line-Plane-Intersection} algorithm and find a new vertex $v$.
Its index $id_v$ will be equal to the maximum value in \textit{polyF} plus one, just to make sure that we are not using the index of an existing vertex.
Now, if $v_1$ is above $p$ (and consequently $v_2$ is below) we only add $v$ to \textit{above}, while if $v_1$ is below $p$ (and $v_2$ is above) we add both $v$ and $v_2$.
As already noted in Section~\ref{subsec:polyhedron_plane}, treating separately the weak intersections makes the code simpler and more efficient.

\begin{algorithm}[htbp]
\caption{Polygon-Plane-Intersection}
\label{alg:polygon-plane}
\begin{algorithmic}[1]
\Require Points \textit{polyV}, Face \textit{polyF}, Plane $p$.
\Ensure Points \textit{aboveV}, Face \textit{aboveF}.
\State evaluate the position of all points in \textit{polyV} with respect to $p$;
\For{$i=1$ : size(\textit{polyF})}
     \State $id_1:=$ \textit{polyF}$(i)$, 
     $id_2:=$ \textit{polyF}$(i+1)$;
     \State $v_1:=$ \textit{polyV}$(id_1)$, 
     $v_2:=$ \textit{polyV}$(id_2)$;
     \If{$v_1$ and $v_2$ are strictly below $p$}
        \State continue;
    \ElsIf{$v_1$ is weakly below $p$ and $v_2$ is on $p$}
        \State \textit{aboveV} $\leftarrow v_2$, \textit{aboveF} $\leftarrow id_2$;
    \ElsIf{$v_1$ and $v_2$ are weakly above $p$}
        \State \textit{aboveV} $\leftarrow v_2$, \textit{aboveF} $\leftarrow id_2$;
    \Else
        \State $v:=$ Line-Plane-Intersection$(v_1,v_2,p)$;
        \State $id_v:=$ max(\textit{polyF})+1;
        \If{$v_1$ is strictly above $p$}
            \State \textit{aboveV} $\leftarrow v$, \textit{aboveF} $\leftarrow id_v$;
        \Else \
            \State \textit{aboveV} $\leftarrow v$, \textit{aboveF} $\leftarrow id_v$;
            \State \textit{aboveV} $\leftarrow v_2$, \textit{aboveF} $\leftarrow id_2$;
        \EndIf
     \EndIf
\EndFor
\State \Return \textit{aboveV}, \textit{aboveF};
\end{algorithmic}
\end{algorithm}


\subsection{Line-Plane-Intersection}
\label{subsec:line_plane}
This last algorithm returns the intersection point between a line, given as a couple of vertices, and a plane.
It is a very simple and well known procedure, and we report it here only for completeness.

The intersection vertex $v$ is defined by a linear combination of $v_1$ and $v_2$, with a coefficient $t$ which may also fall outside the standard range $[0,1]$.
The only difficulty lies in the definition of $t$, which is the negative ratio between two scalar products involving the plane normal $p.n$ and a generic other point on the plane, $p.s$, other than $v_1$ and $v_2$, see Fig~\ref{fig:polygon-line-plane}(b).
If the denominator $D$ vanishes, it means that either the line does not intersect the plane or that the line is contained in it: we treat these exceptions as errors because in Algorithm~\ref{alg:polygon-plane} we only call this algorithm after checking that the edge $(v_1,v_2)$ properly intersects the plane $p$.

\begin{algorithm}[htbp]
\caption{Line-Plane-Intersection}
\label{alg:line-plane}
\begin{algorithmic}[1]
\Require vertices $v_1,v_2$, Plane $p$.
\Ensure vertex $v$.
\State $N:=(p.n)\cdot(v_1-p.s)$;
\State $D:=(p.n)\cdot(v_2-v_1)$;
\If{$D=0$ \ \& \ $N\neq0$} error: no intersections;
\ElsIf{$D=0 \ \& \ N=0$} error: line contained in the plane;
\EndIf
\State $t:=-N/D$;
\State \Return $v:=v_1+t(v_2-v_1)$;
\end{algorithmic}
\end{algorithm}


\subsection{Computational complexity}
\label{subsec:computational_complexity}
Making advantage of the modularity of our algorithms, we can estimate separately the computational cost of each algorithm and then include them into a single formula.

We start with a polyhedron $P$ with $n_v$ vertices and $n_f$ faces.
For Algorithm~\ref{alg:kernel} we need to compute its AABB, which is $O(n_v)$, and then perform Algorithm~\ref{alg:polyhedron-plane} for $n_f$ times, therefore $N_1=n_v+n_f N_2$.

In Algorithm~\ref{alg:polyhedron-plane} we receive as input a polyhedron which is potentially different from $P$, but we empirically measured that the number of vertices and faces remain approximately constant when applying this routine.
We preliminarly compute $n_v$ signed distances; then, only in the cases in which the vertices of a face do not all lie by the same side, run Algorithm~\ref{alg:polygon-plane}.
If only $m_1$ faces needs to be cut (with $m_1<n_f$), the clippping part takes $n_v + m_1 N_3$ operations.
If we have a cap face (which is not always true) with $n_{vc}$ verts, we need to sort them with a QuickSort routine which is on average $O(n_{vc}\log n_{vc})$.
Altogether, Algorithm~\ref{alg:polyhedron-plane} takes $N_2 =n_v + m_1 N_3 + n_{vc}\log n_{vc}$ operations.

Given a face $f_i$ with $n_{vi}$ vertices, in Algorithm~\ref{alg:polygon-plane} we start by computing $n_{vi}$ signed distances.
Then, if only $m_2$ of the edges intersect the plane (where $m_2<n_{vi}$), perform Algorithm~\ref{alg:line-plane}.
This means that $N_3=n_{vi}+m_2 N_4$, and Algorithm~\ref{alg:line-plane} only consists of 4 operations.

Collecting all costs together, we obtain:
\begin{align*}
    N_4 &= 4;\\
    N_3 &= n_{vi}+4m_2;\\
    N_2 &= n_v + m_1 \left(n_{vi}+4m_2\right) + n_{vc}\log n_{vc};\\
    N_1 &= n_v+n_f \left(n_v + m_1 \left(n_{vi}+4m_2\right) + n_{vc}\log n_{vc}\right);\\
    &=n_v+n_f \left(n_v + C\right).
\end{align*}

Let us focus our attention on the term $C$.
Since $m_2\le n_{vi}$ and $n_{vc}$ is the number of vertices of the cap face, we can substitute them with an average $n_a$ of the number of vertices per face and get $C=m_1 n_a + n_a\log n_a$.
For the very majority of the considered models (especially the more complex ones) we can assume that both $m_1$ and $n_a$ are negligible compared to $n_v$: a plane can intersect a polyhedron in a very limited number of its faces, and the average number of vertices per face is significantly smaller than the total number of vertices.
The term $C$ can therefore be included in $n_v$, and as a realistic approximation of the computational cost we get $O(n_v(1+n_f))$.
%
%


\section{Examples and discussions}
\label{sec:examples}
\begin{figure*}[ht!]
\centering
\begin{tabular}{c}
\includegraphics[width=\linewidth]{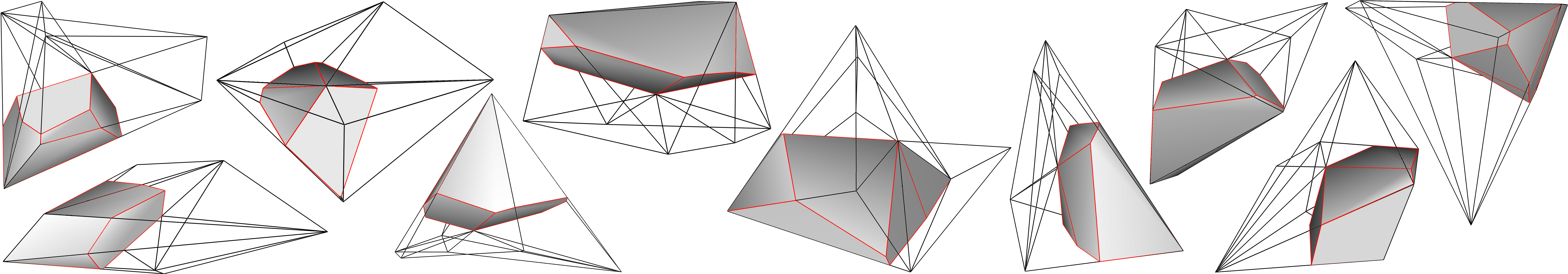} \\
\includegraphics[width=\linewidth]{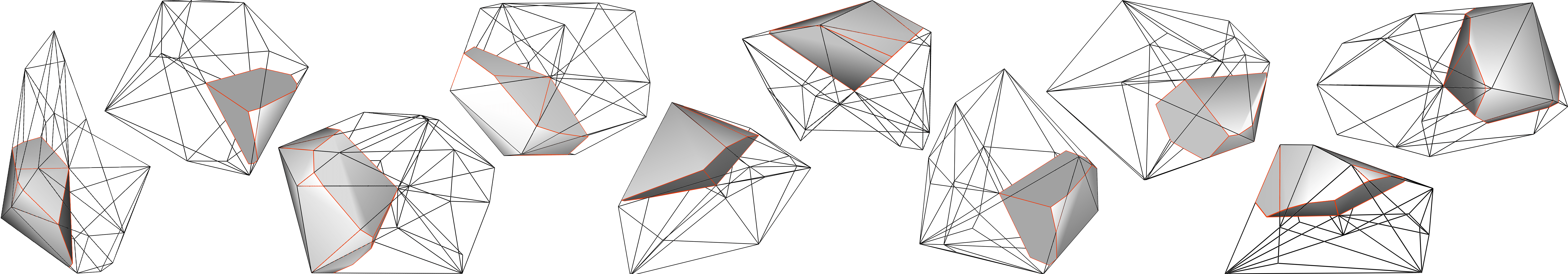} \\
\includegraphics[width=\linewidth]{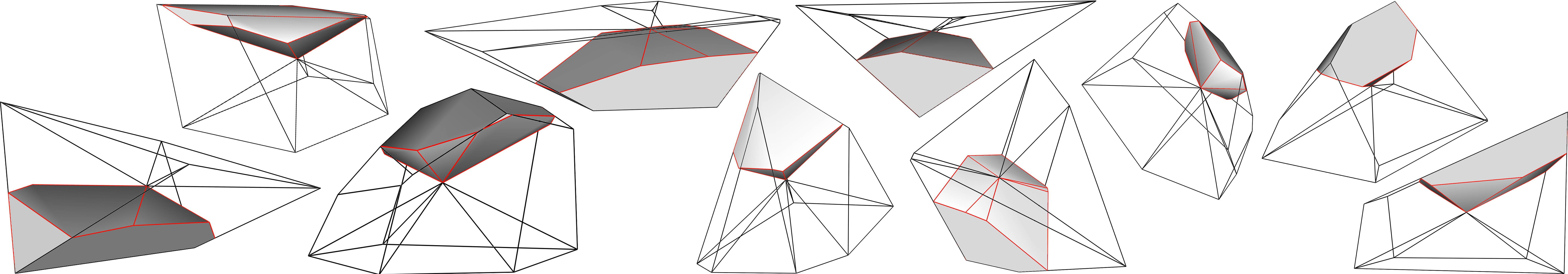}
\end{tabular}
\caption{Polyhedra and kernels from datasets \textit{tet10}, \textit{tet30} and \textit{voro}.}
\label{fig:dataset}
\end{figure*}
In this section we test our method over a collection of polyhedra, comparing its performance to the results obtained 
using our implementation of the algebraic method in CGAL.
Experiments have been performed on a MacBook Pro equipped with a 2,3 GHz Intel Core i5 processor with four CPUs and 16GB of RAM.
Source code is written in C++ and it is accessible at \url{https://github.com/TommasoSorgente/polyhedron_kernel} together with all datasets.

\subsection{Collections of polyhedral elements}
\label{subsec:datasets}
As an initial test we built four datasets containing 1000 non-convex polyhedra each, simulating a typical configuration in which a tessellation of a complex 3D computational domain come with non-convex elements.
As elements of a tessellation on which to perform numerical simulations, they typically present a fairly simple shape and possess a fairly limited number of vertices.
A gallery of examples is visible in Fig.~\ref{fig:dataset}.

The first three are denoted \textit{tet10}, \textit{tet20} and \textit{tet30}, where the number indicates the number of vertices of the contained polyhedra.
Each polyhedron is built by sampling the relative number of points randomly in the space and connecting them in a Delaunay tetrahedalization using \textit{Tetgen} \cite{si2015tetgen}.
Then one vertex is moved to the polyhedron's centroid, in order to get a non convex element.
Note that \textit{Tetgen} outputs the convex hull of the Delaunay tetrahedralization of a set of points, therefore the initial polyhedron is guaranteed to be convex, its centroid is guaranteed to lie inside its volume and the operation of moving a vertex towards the polyhedron centroid does not generate connectivity problems.

The fourth dataset, \textit{voro}, contains 1000 random non-convex polyhedra with non-triangular faces.
For building each polyhedron we start from a Voronoi cell and triangulate the face with the largest area connecting the face vertices to the face centroid.
Then we move the face centroid to the polyhedron centroid to obtain a non convex element.
It is not possible to know a priori the exact number of vertices of each Voronoi cell, but after the definition of dataset \textit{voro} we measured that the number of vertices of its elements ranges from 5 to 20.

\begin{table}[htbp]
\caption{Computational times for estimating the kernel on the whole dataset (in seconds) and ratio between the CGAL time and ours.}
\label{table:time:dataset}
\centering
\begin{tabular}{ccccc}
\hline\noalign{\smallskip}
dataset & $\#$vertices & our & CGAL & ratio\\
\noalign{\smallskip}\hline\noalign{\smallskip}
\textit{tet10} & 10 & 0.29 & 2.49 & 8.58 \\
\textit{tet20} & 20 & 0.49 & 3.62 & 7.39 \\
\textit{tet30} & 30 & 0.63 & 4.49 & 7.13 \\
\textit{voro} & 5-20 & 0.24 & 1.92 & 8 \\
\noalign{\smallskip}\hline
\end{tabular}
\end{table}

We measure the computational time needed to estimate the kernels of all polyhedra in each dataset and compare it with the performance of the CGAL library in Table~\ref{table:time:dataset}.
On this kind of polyhedra, the geometrical approach performs significantly better (between seven and eight times) than the algebraic one.
This may be due to the fact that, with a limited number of planes, computing the geometric intersection between them is cheaper than solving a linear problem.
As the number of vertices increases the difference between the two approaches becomes smaller.

\subsection{Refinements}
\label{subsec:refinements}
As a second setting for our tests we wanted to measure the asymptotic behaviour of our method as the number of vertices increases.
We considered two polyhedra taken from the dataset \textit{Thingi10K} \cite{zhou2016thingi10k}: the so-called \textit{laser-chess} and \textit{flex}.
These models are given in the form of a surface mesh and we treat them as single volumetric cell, analyzing the performance of our algorithm as we refine them.

\begin{figure}[th]
\centering
    \includegraphics[width=.8\linewidth]{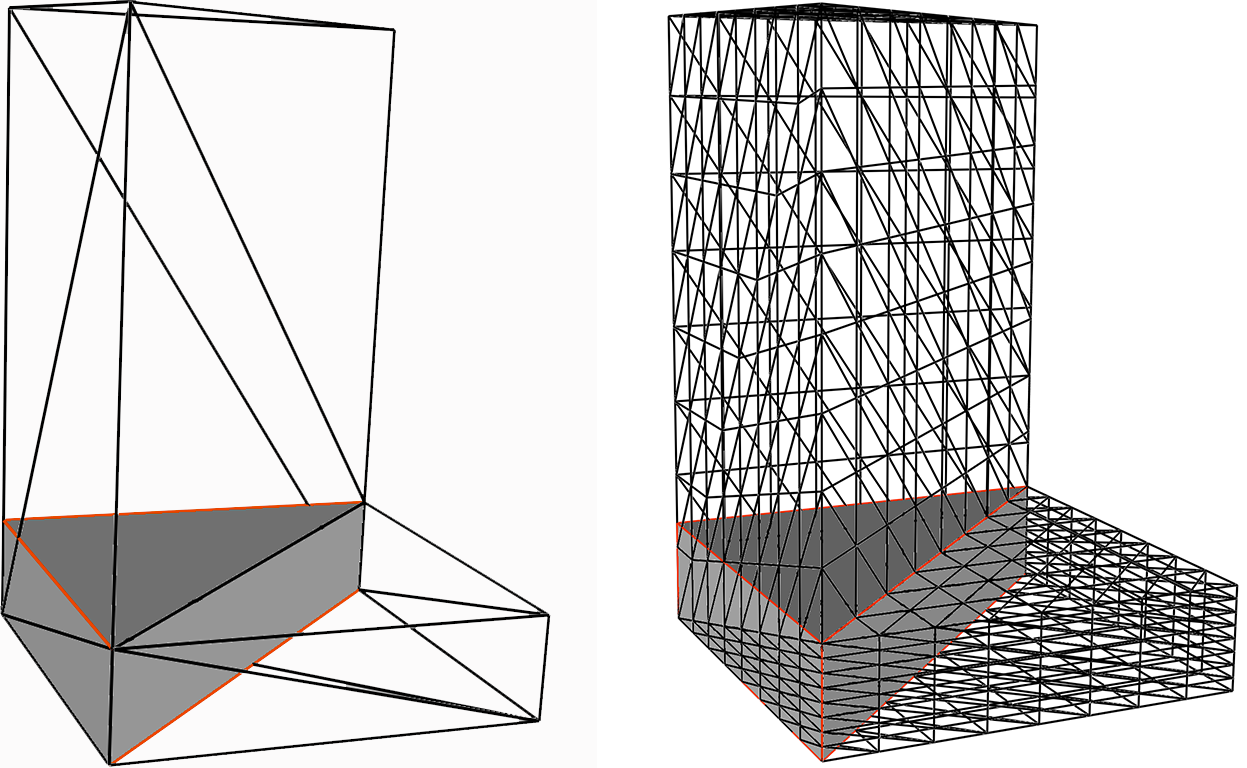}
\caption{Original laser-chess model and its third refinement. The kernels are identical.}
\label{fig:laser-chess}
\end{figure}

\begin{figure}[htbp]
\centering
    \includegraphics[width=.8\linewidth]{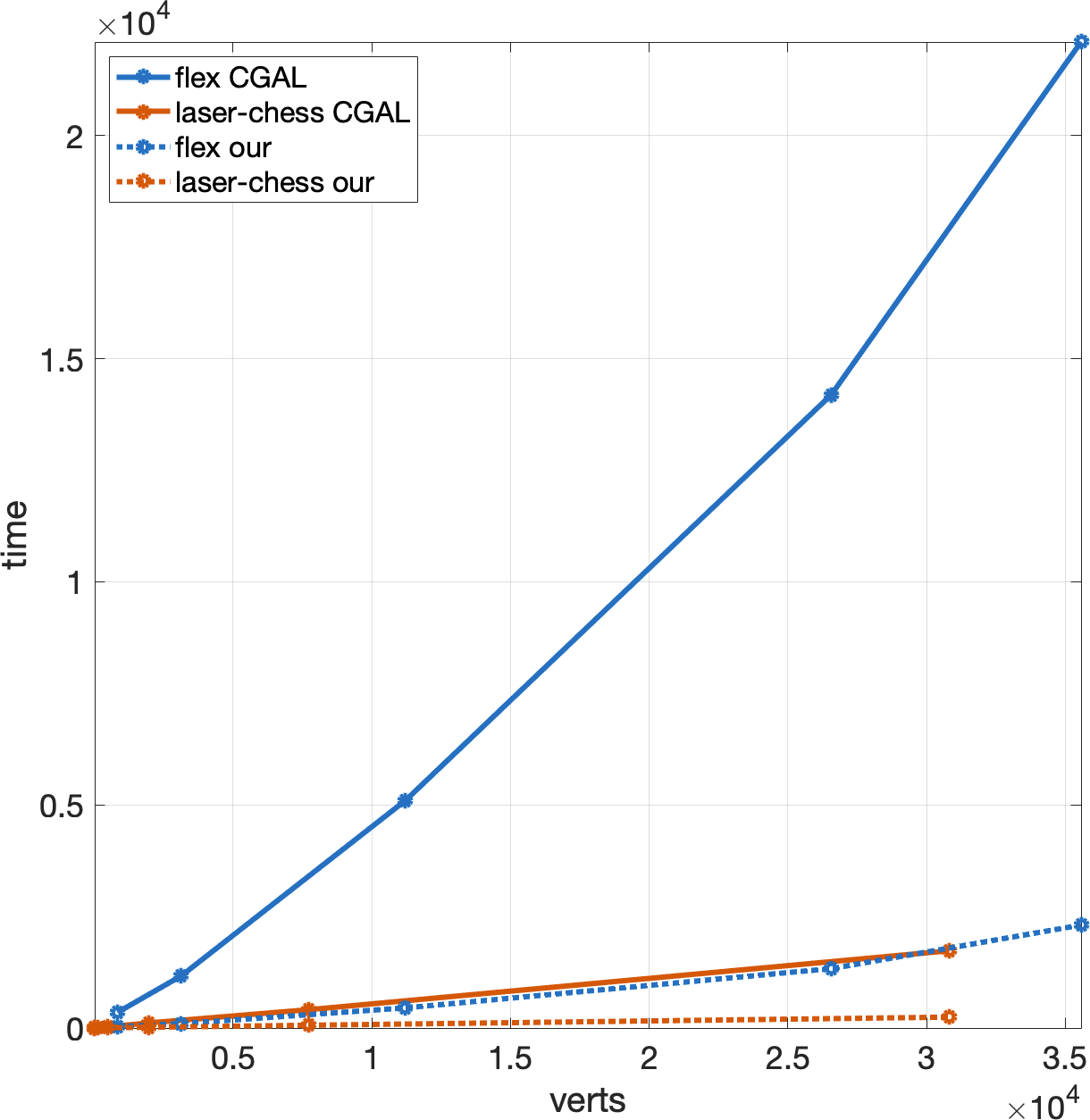}
\caption{Computational times for the refinement of the flex and laser-chess models. Time is expressed in milliseconds.}
\label{fig:rate}
\end{figure}

The \textit{laser-chess} model is extremely simple, being defined by only eight planes.
Even refining its faces with a midpoint strategy, the planes induced by its faces remain identical and the kernels of the refined models are all equal (Fig.~\ref{fig:laser-chess}).
On this example our method performs on average 6.8 times better that the algebraic method (see Table~\ref{table:time:refinements}), and the computational time scales with a constant rate (see the red lines in Fig.~\ref{fig:rate}).
Our implementation takes advantage of the fact that Algorithm~\ref{alg:polyhedron-plane} recognises coplanar faces and performs Algorithm~\ref{alg:polygon-plane} only eight times, independently of the number of faces.

\begin{figure}[htbp]
\centering
    \includegraphics[width=.95\linewidth]{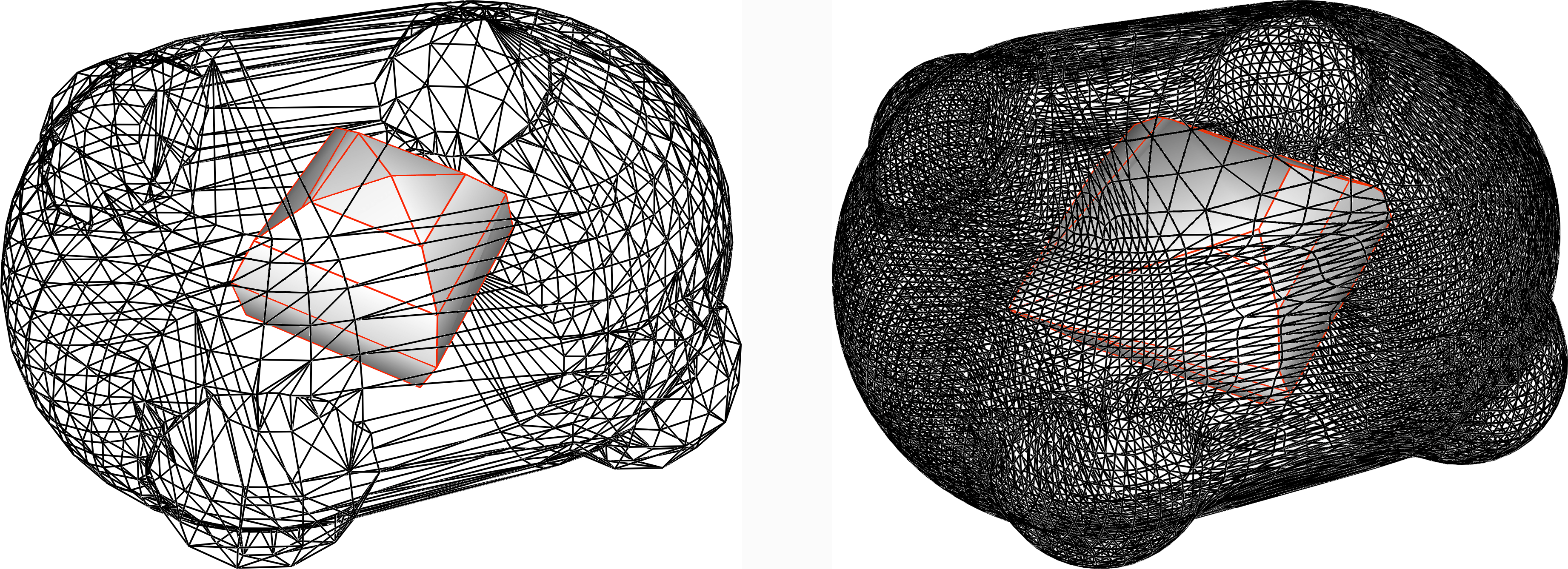}
\caption{Original flex model and its third refinement. Small perturbations in the faces lead to slightly different kernels.}
\label{fig:flex}
\end{figure}

The \textit{flex} model is more complex, as it presents a curved surface which generates a lot of different planes defining the kernel.
Moreover, we refined this model using the Loop's algorithm and this generated faces lying on completely new planes.
This explains the difference between the two kernels in Fig.~\ref{fig:flex}.
Our geometric method improves the performance of the algebraic one by one order of magnitude, even if the difference between the two approaches decreases (from 13 times to 9, see Table.~\ref{table:time:refinements}) as the number of vertices increases.

\begin{table}[htbp]
\caption{Computational times for the laser-chess and flex refinements (in seconds) and ratio between the CGAL time and ours.}
\label{table:time:refinements}
\centering
\begin{tabular}{ccccc}
\hline\noalign{\smallskip}
mesh & $\#$vertices & our & CGAL & ratio \\
\noalign{\smallskip}\hline\noalign{\smallskip}
\textit{laser-chess1} & 10 & 0 & 0.001 & 1 \\
\textit{laser-chess2} & 127 & 0.001 & 0.007 & 7 \\
\textit{laser-chess3} & 493 & 0.004 & 0.027 & 6.75 \\
\textit{laser-chess4} & 1945 & 0.017 & 0.11 & 6.47 \\
\textit{laser-chess5} & 7729 & 0.06 & 0.41 & 6.83 \\
\textit{laser-chess6} & 30817 & 0.24 & 1.73 & 7.21 \\
\hline
\textit{flex1} & 834 & 0.026 & 0.35 & 13.46 \\
\textit{flex2} & 3130 & 0.097 & 1.17 & 12.06 \\
\textit{flex3} & 11216 & 0.45 & 5.1 & 11.33 \\
\textit{flex4} & 26560 & 1.33 & 14.19 & 10.67 \\
\textit{flex5} & 35566 & 2.31 & 22.1 & 9.57 \\
\textit{flex6} & 42659 & 3.57 & 33.26 & 9.32 \\
\noalign{\smallskip}\hline
\end{tabular}
\end{table}

\subsection{Complex models}
\label{subsec:complex_models}
\begin{figure*}[ht]
\centering
    \includegraphics[width=\linewidth]{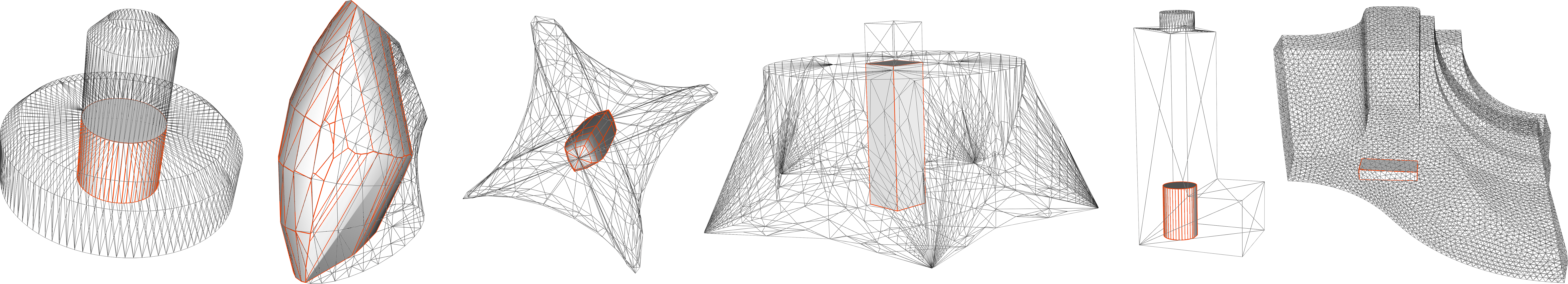}
\caption{Examples of our kernel evaluation for complex models: \textit{bot-eye, rt4-arm, super-ellipse, playset, leg, fandisk}.}
\label{fig:complex}
\end{figure*}
Last, we try to compute the kernel of some more complex models, taken again from the dataset \textit{Thingi10K} and treated as single volumetric cells.
Note that most of the models in Thingi10K are not star-shaped, thus making it useless to compute their kernels.
Even if our method is designed for dealing with a large number of simple polyhedra (i.e., with a relatively small number of vertices), our algorithms are still able to compute the kernel of a whole object with thousands of vertices.
In Fig.~\ref{fig:complex} we present the kernel computation of six complex objects, and the computational times are reported in Table~\ref{table:time:complex}.

\begin{table}[th]
\caption{Computational times for complex models (in seconds) and ratio between the CGAL time and ours.}
\label{table:time:complex}
\centering
\begin{tabular}{ccccc}
\hline\noalign{\smallskip}
mesh & $\#$vertices & our & CGAL & ratio \\
\noalign{\smallskip}\hline\noalign{\smallskip}
\textit{bot-eye} & 453 & 0.28 & 0.048 & 0.17\\
\textit{rt4-arm} & 655 & 0.21 & 0.11 & 0.52\\
\textit{super-ellipse} & 290 & 0.02 & 0.04 & 2\\
\textit{playset} & 1416 & 0.01 & 0.08 & 8\\
\textit{leg} & 87 & 0.003 & 0.03 & 10\\
\textit{fandisk} & 7229 & 0.07 & 3.58 & 51.14\\
\noalign{\smallskip}\hline
\end{tabular}
\end{table}

We notice that the number of vertices of the element, by itself, is not strictly related to the efficiency of the method.
Much more influence has the shape of the object: in accordance to the results of Section~\ref{subsec:refinements}, over simple models like \textit{leg} or models with numerous ajacent coplanar faces like \textit{playset} and \textit{fandisk} our method is preferable.
Note in particular the different performance of the two methods over the \textit{fandisk} model, which has a high number of vertices.
Vice versa, over elements with significant curvatures like \textit{bot-eye}, \textit{rt4-arm} or \textit{super-ellipse}, the algebraic method performs similarly or better than ours even on relatively small models.

\section{Conclusions}
\label{sec:conclusions}
We presented an algorithm for the computation of the kernel of a polyhedron based on the extension to the 3D case of the geometric approach adopted in two dimensions.
The algorithm showed up to be robust and reliable, as it computed successfully the kernel of every considered polyhedron.

We compared its efficiency to the one of the algebraic approach to the problem, implemented in CGAL.
From a theoretical point of view, the computational complexity evaluation of Section~\ref{subsec:computational_complexity} suggests that our method is in general quadratic, while the algebraic approach has a lower bound at $n\log(n)$.
Nonetheless, we proved in Section~\ref{sec:examples} that in several circumstances our approach outperforms the algebraic one.

Our method performs significantly better than the algebraic approach over polyhedra with a limited number of vertices and faces, as shown in Section~\ref{subsec:datasets}, making it particularly suitable for the analysis of volumetric tessellations with non-convex elements. Indeed, we point out that our algorithm is specifically designed to be used with simple polyhedra, possibly composing a bigger and more complex 3D model, and not with a complete model itself.

When the size of the polyhedron increases, our method is still particularly efficient if the model has numerous coplanar faces, like in some of the complex examples in Section~\ref{subsec:complex_models}.
This is a very common situation in models representing mechanical parts.
On the other side the algebraic approach is preferable over curved domains, with numerous vertices and faces lying on different planes.

In conclusion, with this work we do not aim at completely replacing the algebraic approach for the kernel computation but instead to give an alternative which can be preferred for specific cases, such as the quality analysis of the elements in a 3D tessellation, in the same way as bubble-sort is to be preferred to optimal sorting algorithms when dealing with very small arrays.

As a future development, we plan to include this tool in a suite for the generation and analysis of tessellations of three dimensional domains, aimed at PDE simulations.
The kernel of a polyhedron has a great impact on its geometrical quality, and the geometrical quality of the elements of a mesh determines the accuracy and the efficiency of a numerical method over it.
We therefore plan to use this algorithm for better understanding the correlations between the shape of the elements and the performance of the numerical simulations, and be able to adaptively generate, refine or fix a tessellation accordingly to them.

\section*{Acknowledgements}
We would like to thank Dr. M. Manzini for the precious discussions and suggestions, and all the people from IMATI institute involved in the CHANGE project.
Special thanks are also given to the anonymous reviewers for their comments and suggestions.

This work has been partially supported by the ERC Advanced Grant CHANGE contract N.694515.

\bibliographystyle{plain}

\end{document}